\font\bigboldiii=cmbx10 scaled\magstep3
\def\UCRNOBIT{
{{\it                   Department of Physics\break
University of California at Riverside\break
                  Riverside, California 92521--0413; U.S.A.}}}
\def\noblackbox{\overfullrule=0pt}
\def\sm{Standard Model}
\def\gesim{\,{\raise-3pt\hbox{$\sim$}}\!\!\!\!\!{\raise2pt\hbox{$>$}}\,}
\def\lesim{\,{\raise-3pt\hbox{$\sim$}}\!\!\!\!\!{\raise2pt\hbox{$<$}}\,}
\def\su#1{{SU(#1)}}
\def\gev{~\hbox{GeV}}
\def\tev{~\hbox{TeV}}
\def\lowti#1{_{{\rm #1 }}}
\def\inv#1{{1\over#1}}
\def\sqr#1#2{{\vcenter{\hrule height.#2pt
      \hbox{\vrule width.#2pt height#1pt \kern#1pt
         \vrule width.#2pt}
      \hrule height.#2pt}}}
\def\square{\mathchoice\sqr56\sqr56\sqr{2.1}3\sqr{1.5}3}

\input epsf
\def\thetitle{The meaning of anomalous couplings}
\def\theabstract{A prescription is presented for the interpretation of
the coefficients in an effective lagrangian in terms of physical mass
scales.}
\def\ucrnum{164}
\def\theauthor{\author{{\fourteencp J.
Wudka\foot{jose.wudka@ucr.edu}}}\address{\UCRNOBIT}}

\input phyzzx
\Twelvepoint
\PHYSREV
\def\square{\mathchoice\sqr56\sqr56\sqr{2.1}3\sqr{1.5}3}

\rightline{UCRHEP-T\ucrnum}
{\titlepage
\vskip -.2 in
\title{ {\bigboldiii \thetitle}}
\doublespace
\author{\theauthor}
\abstract
\bigskip
\singlespace
\theabstract
\endpage}

\REF\reviews{See, for example, the reviews,
H. Georgi, {\it Ann. Rev. Nucl. Part. Science} {\bf43}, 209 (1994).
J. Polchinski,  Lectures presented at {\sl TASI 92}, Boulder, CO, Jun 3-28,
1992.
A. Pich, Lectures given at {\sl 5th Mexican School of Particles and Fields},
Guanajuato, Mexico, 30 Nov - 11 Dec 1992.
G. Ecker, {\it Prog. Part. Nucl. Phys.} {\bf35}, 1 (1995).
J. Wudka, {\it Int. J. Mod. Phys.} {\bf A9}, 2301 (1994).}

\REF\example{
K. Hagiwara \etal, {\it Nucl.Phys.}, {\bf B282}, 253 (1987).
G.L. Kane \etal, {\it Phys. Rev.} {\bf D39}, 2617 (1989).
U. Baur and E.L. Berger (Argonne), {\it  Phys. Rev.} {\bf D41} 1476 (1990).
T. Helbig and H. Spiesberger, {\it Nucl. Phys.} {\bf B373}, 73 (1992).
X.-G. He and B. McKellar, {\it Phys. Lett.} {\bf B320}, 165 (1994).
}

\REF\yao{
G.-L. Lin \etal, {\it Phys. Rev.} {\bf D49}, 2414 (1994); {\it ibid} {\bf D44},
2139 (1991).
H. Steger \etal, {\it  Phys. Rev. Lett.} {\bf59}, 385 (1987).}

\REF\brook{V.W. Hughes, AIP conference proceedings no. 187, (1989) 326.  M.
May, AIP conference proceedings no. 176, (1988) 1168.}

\REF\marty{M.B. Einhorn, {\it Phys. Rev.} {\bf D49}, 1668 (1994).}

\REF\gmtwo{C. Arzt. \etal, {\it  Phys. Rev.} {\bf D49}, 1370 (1994).}

\REF\veltman{M. Veltman, {\it Acta Phys. Polon.} {\bf B12}, 437 (1981).}

\REF\tree{C. Arzt. \etal, {\it Nucl. Phys.} {\bf B433}, 41 (1995).}

\REF\bw{
C.J.C. Burges and H.J. Schnitzer, {\it Nucl. Phys.} {\bf B228}, 464 (1983).
C.N. Leung \etal, {\it Z. Phys.} {\bf C31}, 433 (1986).
W. Buchm\"uller and D. Wyler, {\it Nucl. Phys.} {\bf B268}, 621 (1986).}

\REF\bandl{C.P. Burgess and D. London, report MCGILL-92-04. e-Print Archive:
hep-ph/9203215.}

\REF\zzz{
U. Baur and D. Zeppenfeld, {\it Phys. Lett.} {\bf201B}, 383 (1988).
U. Baur and E.L. Berger, {\it Phys. Rev.} {\bf D47}, 4889 (1993).}

\REF\agbi{H. Aihara \etal, Summary of the Working
Subgroup on Anomalous Gauge Boson Interactions of the DPF Long-Range Planning
Study, to be
published in {\sl Electroweak Symmetry Breaking and Beyond the Standard Model},
eds. T. Barklow,
S. Dawson, H. Haber and J. Seigrist. e-Print Archive: hep-ph/9503425.}

\REF\okun{
L.B. Okun, {\sl Leptons and quarks}, sect. 6.4 (North Holland, 1984).
Review of Particle Properties, {\it Phys. Rev.} {\bf D50}, 1177 (1994).}

\REF\gw{B. Grzadkowski and J. Wudka, {\it Phys. Lett.} {\bf B364}, 49 (1995).}

\REF\ivf{A. Djouadi \etal, in proceeding of the {\sl Workshop on Physics
and Experiments with Linear Colliders}, Saariselka, Finland, Sep 9-14,
1991.}

\REF\ehlq{See, for example, E. Eichten \etal, {\it Rev. Mod. Phys} {\bf56}, 579
(1984), ERRATUM-{\it ibid} {\bf58}, 1065 (1986).}

\REF\nda{
S. Weinberg, {\it Physica} {\bf 96A}, 327 (1979).
H. Georgi and A. Manohar, {\it  Nucl. Phys.} {\bf B234}, 189 (1984).
H. Georgi, {\it Phys. Lett.} {\bf B298}, 187 (1993).}

\REF\nat{G. 't Hooft, lecture given at {\sl Cargese Summer Inst.}, Cargese,
France, Aug 26 - Sep 8, 1979.}

\REF\bound{H. Georgi and A. Manohar, {\it  Nucl. Phys.} {\bf B234}, 189 (1984).
T. Appelquist and  G.-H. Wu, {\it Phys. Rev.} {\bf D48}, 3235 (1993).}

\REF\gl{
J. Gasser and H. Leutwyler, {\it Nucl. Phys.} {\bf B250}, 465 (1985).
R.S. Chivukula \etal, {\it  Ann. Rev. Nucl. Part. Sci.} {\bf45}, 255 (1995).}

\REF\zbb{
E. Ma \etal, {\it Phys. Rev.} {\bf D53}, 2276 (1996). 
G. Bhattacharyya \etal, report CERN-TH-95-326 (unpublished). e-Print Archive:
hep-ph/9512239.
T. Yoshikawa \etal, report HUPD-9528 (unpublished). e-Print Archive:
hep-ph/9512251.
G. Altarelli \etal, report CERN-TH-96-20 (unpublished). e-Print Archive:
hep-ph/9601324.
C.-H. V. Chang \etal, report NHCU-HEP-96-1 (unpublished). e-Print Archive:
hep-ph/9601326.
P. Bamert \etal, report MCGILL-96-04 (unpublished). e-Print Archive:
hep-ph/9602438.
K.S. Babu \etal, report IASSNS-HEP-96-20 (unpublished). e-Print Archive:
hep-ph/9603212.
D. Comelli and J.P. Silva, report HEPPH-9603221 (unpublished). e-Print Archive:
hep-ph/9603221.}

\noblackbox

\advance \hsize by 1.5 in
\advance \hoffset by -0.5 in
\advance \vsize by .8 in
\advance \voffset by -0.6 in

\baselineskip 17 pt

\chapter{Introduction}

During the last 15 years many papers have appeared which attempt
to describe physics beyond the Standard Model in a model
independent way using effective lagrangians~\refmark{\reviews}. Many of these
papers present scenarios where enormous deviations from the Standard
Model~\refmark{\example} are obtained. Given these situations one
might reasonably ask whether there
is {\it any} model that could generate such striking deviations,
and, even in case where no such deviations are observed, what
constraints on the underlying theory can be inferred from the experimental
limits.

In this short note I want to describe how one can answer these
questions. Since we have not observed yet any clear deviation from
the Standard Model, the constraints on new physics are not
completely unambiguous. Yet there are several statements that
can be made irrespective of the kind of new physics which awaits
us. The aim is to provide a sound recipe for extracting limits on the
scale of new physics from the experimental bounds on the deviations from
the Standard Model.

I will first motivate the results using electroweak physics as an
example. Then I will discuss weakly and strongly interacting heavy
physics concentrating on the interesting case of the vector-boson
interactions.

Most of the contents in this paper have appeared in various
publications; my purpose is to present a summary of the
results.

\chapter{Electroweak interactions as an example}

When considering the low-energy limit of a given theory one has
(inevitably) to deal with effective interactions produced by virtual
heavy physics effects~\refmark{\reviews}.
Thus, when we consider the low-energy limit of
the electroweak sector~\refmark{\yao} of the Standard Model, one obtains
Fermi's
theory of the weak interactions. QED for all light fermions is
also generated, together with a host
of other interactions such as those describing the weak contributions
to the fermion's anomalous magnetic moments, the $W$ and heavy quark
contributions to the Euler-Heisenberg lagrangian, etc.

All these non-renormalizable interactions come with dimensional
coefficients. For example, the term describing the
weak contributions to the
anomalous magnetic moment
of the muon, $$ \bar \mu \sigma^{ \alpha \beta } \mu \, F_{ \alpha
\beta}, \eqn\eq$$ has dimension five and will appear multiplied by
a constant of mass dimension $-1$. The four-fermion operators
describing low-energy electroweak physics have dimension six and
appear multiplied by a constant of mass dimension $-2$, etc.

When considering a specific theory such as the Standard Model all these
dimensional coefficients can be calculated in terms of the gauge coupling
constants, the $W$ and $Z$ masses, etc. Such interactions inherit the
symmetries of the \sm; not all possible Lorentz invariant terms occur.

With the benefit of hindsight, we may reinterpret the history of the
\sm: when we were ignorant about the details
of the electroweak theory what we did was write down the most general
set of operators which

\item{{\it(i)}} Contained light fields (eg. leptons).

\item{{\it(ii)}} Respected the QED gauge symmetry (but not
necessarily its global symmetries such as P and C!)

\smallskip

Such operators appeared multiplied by unknown coefficients which were
constrained by the data. In this way we realized that the charged-current
couplings
were of the V--A type and not something else. We were also able to
get an idea of the order of magnitude of the scale of the new physics,
and for this we used the fact that the four-fermion operators could
be generated by some heavy particle through a process such as the
one in the figure below.

\setbox2=\vbox to 150pt{\epsfysize=6 truein\epsfbox[0 -100 612
692]{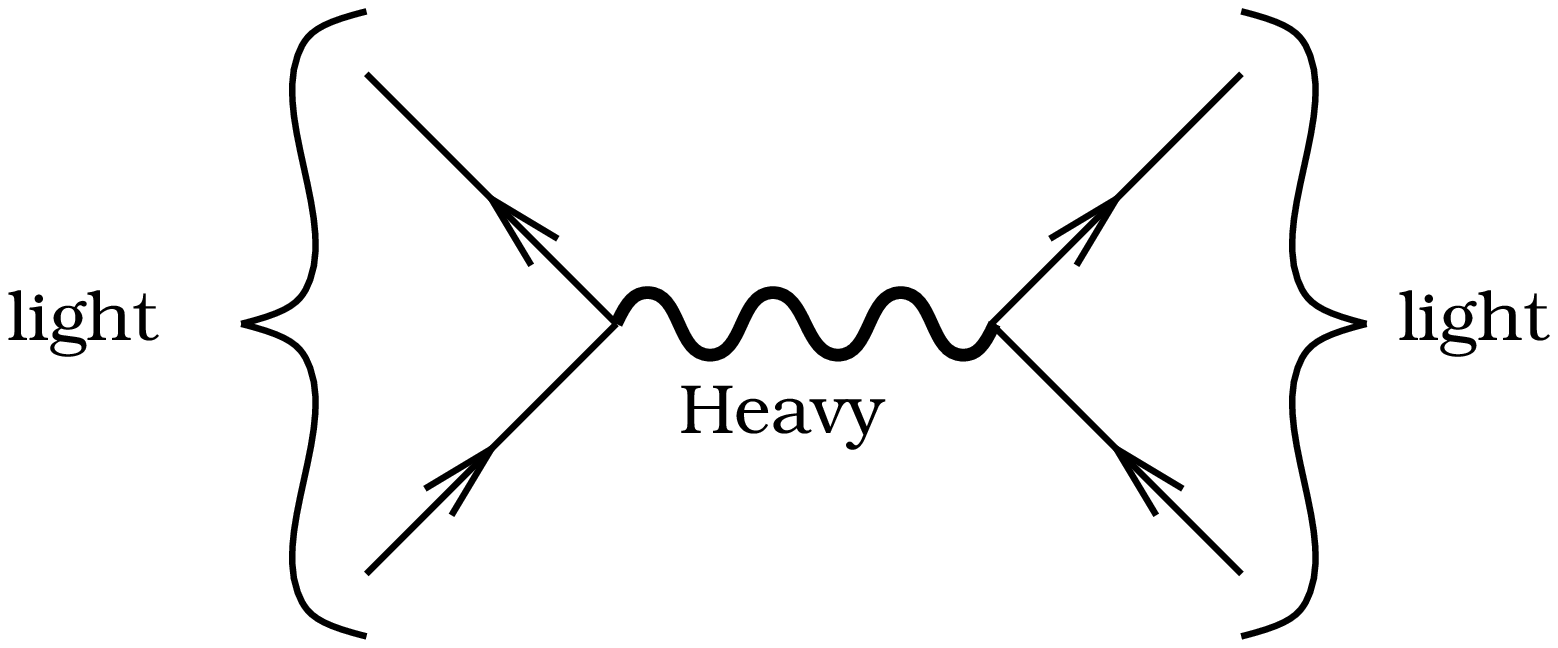}}
\centerline{\box2}

{}From such ideas we concluded that the coefficient of the four-fermion
interactions would be of the form $$ \left(
\hbox{ coupling constant } \over \hbox{ heavy
mass } \right)^2. \eqn\eq$$ If we assume that the theory underlying the weak
interactions is weakly coupled, so that $($coupling constant $) \lesim
1 $, we could get an estimate of the heavy mass from the observation
of the processes mediated by the four-fermion interactions. We
then designed colliders to probe physics at that energy.

Note also the two following points

\item{\bullet} We did {\it not} expect the four-fermion interactions to be an
accurate description of weak processes at energies close to the scale we
just inferred. For example, the four-fermion theory
cannot describe weak physics near the $Z$ pole.

\item{\bullet} Not all weak effects are so amenable to observation. For
example,
the weak contributions to the anomalous magnetic moments are very small
(at the $ 10^{-9} $ level) and only now are being
probed at the Brookhaven experiment AGS 821~\refmark{\brook}~\foot{
Extracting these effects from the data presents other problems,
see~\refmark{\marty} and references therein.}.
This is not because these weak contributions are accidentally suppressed,
nor are they forbidden by some symmetry. The reason is simply that
they are generated by loops, and thus are naturally small~\refmark{\gmtwo}.

Thus when studying the weak interactions we took a very sensible approach:
we selected those effects which are tree-level generated, and
then restricted our interest to those processes forbidden by QED.
By doing this we optimized the chances of obtaining information about
the interactions ultimately responsible for Fermi's effective theory.
This is, of course, an unfair oversimplification of the history, but it
does emphasize the following points

\item{{\bf(a)}} When we want to discover new physics through the virtual
processes which it induces, the most sensible thing to do is to select the
effects that can be generated at tree level, and leave the study of
loop-generated effects for later\foot{Though this is appropriate for
weakly coupled theories similar considerations apply for strongly coupled
ones, see below.}.

\item{{\bf(b)}} When describing the underlying theory through its low-energy
manifestations, one should have an idea of which processes are responsible for
the effective operators we are considering (eg. a $Z$ mediating $ e^+ e^-
\rightarrow \nu \bar \nu $). In this way we can obtain rough estimates
of the {\bf physical} scales involved in the theory.

\item{{\bf(c)}} A description of new physics in terms of effective
operators has a limited range of validity. One can derive from the
formalism the scale at which such description will fail~\refmark{\reviews}.
Applying the formalism beyond such scale will give {\bf wrong} results.

\smallskip

Even if we had been unlucky and found no deviations from QED. The above
process would have provided a lower bound for the scale of new physics.

These considerations, though somewhat self-evident, are regularly ignored
when considering physics beyond the Standard Model through effective
operators. It is not true that ``anything can
happen beyond the Standard Model'':
the fact that the Standard Model is so well
measured puts very strong restrictions on the kinds of new physics that
could be responsible for the virtual effects we are attempting to
measure. This is true even if we have not probed all possible processes:
a strong deviation from the \sm\ in, for example, the $WWZ$ coupling
cannot occur in isolation, a host of other effects must be present
concurrently which are constrained by current data.

The main restriction on the underlying theory is that it should respect
the Standard Model gauge symmetries. If we assume that this is not the
case then, even if the deviations from gauge invariance occur at a scale
$ \Lambda $, they will induce gauge variant terms to which existing
data is sensitive~\refmark{\veltman}. In this case the many
consequences of gauge invariance, such as lepton universality, would be
nothing more than amazing coincidences. It is also important to note that
in {\it all} models studied to date low energy gauge invariance is respected
by the underlying theory.

These restrictions do not extend to global symmetries. For example
$\su5$ GUT does respect the Standard Model gauge symmetry but violates
lepton and baryon numbers.

\chapter{Weakly coupled theories}

If the physics beyond the \sm\ were weakly coupled,
 and if we are interested in
the virtual effects from the heavy particles, the relevant question is
which of the manifold of terms generated at low energies could
possibly be generated through tree-level graphs in the heavy theory.
The list of such tree-induced processes is known~\refmark{\tree}
 (provided we make the
single mild assumption that the underlying theory is a gauge theory); the
list of {\it all} terms of dimension 6 is also known~\refmark{\bw}

All loop-generated terms are small, not necessarily unobservable, but
harder to deal with. A reasonable strategy is to concentrate first on
the observables that can get large contributions from the heavy physics.

To illustrate the consequences of these statements I will consider the
possible modifications to the gauge-boson couplings induced by the heavy
interactions.

\section{Vector-boson interactions.}

Any interaction among vector bosons
not contained in the Standard Model appears as an operator of dimension
six or higher~\refmark{\bw}. In unitary gauge such an operator might appear to
have
dimension 4, but this is because in this gauge (and ignoring Higgs
interactions) the scalar doublet is replaced by a number ($=246\gev$).

All dimension 6 operators mediating vector-boson interactions
are generated by loops, operators of
dimension 8 and higher can be generated at tree level. Again I
emphasize that such operators may appear as if they were dimension 4 operators
in some gauges, but fundamentally they are not.

Because of their origin such operators get a coefficient
$$ \hbox{ dimension 6} : \quad
\sim \inv{ 16 \pi^2 \Lambda^2 } \qquad\qquad
\hbox{ dimension 8} : \quad
\sim \left( { v \over \Lambda } \right)^4 , \eqn\eq$$
where $ \Lambda $ denotes a {\it physical} mass scale, \ie, the mass of
a particle or other similar threshold.

The standard notation~\refmark{\example} is not derived from the effective
lagrangian
approach based on gauge invariant operators, but uses an effective
Lagrangian restricted only through Lorentz and QED gauge invariances.
Nonetheless the arguments described above can be used to interpret the
couplings which appear in the standard approach.

Consider for example the $WWZ$ interaction~\refmark{\example}
$$ - i e \cot \theta_w { \lambda
\over M_w^2} W^+_{ \alpha \beta } W^-_{\beta \mu} Z_{ \mu \alpha } \eqn\eq$$
where $e$ denotes the proton charge, $ \theta_w$ the weak-mixing angle,
and $ W^\pm_{ \alpha \beta } = \partial_\alpha W^\pm_\beta -
\partial_\beta W^\pm_\alpha , \; Z_{ \alpha \beta } = \partial_\alpha Z_\beta -
\partial_\beta Z_\alpha $ are the field strengths for the $W$ and $Z$
vector-boson fields. The coupling $ \lambda $ is unknown and
parametrizes a certain type of new physics;
the gauge invariant formalism provides the estimates~\refmark{\gmtwo}
$$ \lambda \sim
\cases{ {6 M_w^2 g^2 \over 16 \pi^2 \Lambda^2} & for dim-6 operators , \cr
{ M_w^4 \over \Lambda^4} & for dim-8 operators , \cr}
\eqn\estimatesi $$ where, as before, $ \Lambda $
denotes the mass of a heavy excitation, $g$ is the $\su2$ gauge
coupling constant and $ M_w$ the $W$ mass (the factor of 6 is
due to combinatorics).

Using the above estimate one can understand what the experimental
limits imply with respect to the underlying theory (corresponding
to dimension 6 and 8 terms respectively) $$ \lambda \sim \inv{ (
100 \Lambda\lowti{TeV} )^2 } , \quad \inv{ (
12.5 \Lambda\lowti{TeV} )^4 } \eqn\eq$$ where $ \Lambda\lowti{TeV} $ denotes
the scale of new physics in \tev\ units. Thus the statement $ | \lambda
| < 0.1 $ corresponds to $ \Lambda > 150 \gev $ while $ | \lambda | <
10^{-4} $ implies $ \Lambda > 1 \tev $.

This illustrates the power of the gauge invariant approach: we are able
to interpret the results in terms of physical quantities and determine
the implications on the scale of new physics~\foot{It has been shown
argued~\refmark{\bandl} that any theory can be rendered gauge invariant
by introducing spurious degrees of freedom. In this approach, however,
all fermions (and scalars, when present) are assumed to be gauge singlets;
nor is the gauge group uniquely fixed. Taking the \sm\ as an excellent
approximation to the low energy physics excludes this approach; for a
discussion see J. Wudka in Ref. \reviews.}.
For the case of the $WWZ$
interactions if we wish to probe physics at the $ 1 \tev $ scale we must
be able to measure $ \lambda $ to to a precision of $ \sim 10^{-4} $.
This should be done, of course, at colliders whose CM energies lie
below $ 1 \tev $, otherwise
the heavy particles would be produced directly and the effective
operator formalism fails, just as the four-fermion theory should not be
used at energies $ \gesim 80 \gev $.

Similar considerations apply, for example, to the $Z\gamma\gamma $ couplings.
Here it is
known~\refmark{\bw} that the operators responsible for such couplings are
of dimension 8 or higher and can be generated via tree
graphs~\refmark{\tree}. The
coefficients are then expected to be of the form $ 1/\Lambda^4 $. The standard
notation for this case~\refmark{\zzz} is based, again, on a lagrangian
restricted only
by Lorentz and QED gauge invariances. As an example consider the
interactions $$ i { h_3^Z \over M_z^2} \left[ ( \square
+ M_z^2 ) Z_\mu \right]
\tilde F^{ \mu \nu } Z^\nu , \qquad
i { h_4^Z \over M_z^4} \left[ ( \square + M_z^2 ) \partial_\alpha Z_{
\mu \nu } \right] \tilde F^{ \mu \nu } Z^\alpha , \eqn\eq$$ denoting
by $Z_\alpha$ the
$Z$-boson field, $Z_{\mu\nu} = \partial_\mu Z_\nu - \partial_\nu Z_\mu$
 and $ \tilde F_{\mu \nu }$ is the dual of the photon field strength.
We then have the estimates $$ h_3^Z \sim {
v^2 M_z^2 \over \Lambda^4 } \simeq \inv{ \left( 6.7 \Lambda\lowti{TeV}
\right)^4 } , \qquad h_4^Z \sim { M_z^4 \over
\Lambda^4 } \simeq \inv{ \left( 11 \Lambda\lowti{TeV}
\right)^4 } , \eqn\estimatesii $$ so that a bound $ | h_{3,4}^Z | < 1 $ implies
$ \Lambda
> 150 \gev $ and $ \Lambda > 92 \gev $ respectively.

\subsection{Form factors}

It has been customary to use
form factors to insure the theory does not violate unitarity. I will not
do this here for the following reasons:

\item{(i)} The effective lagrangian approach should not be extended to
scales close to a threshold. All attempts at modifying the formalism to
this end are {\it extremely} model dependent and no general conclusions
can be derived from them.

\item{(ii)} The form factors are usually chosen so that there are no poles in
{\it any} physical process. This is unreasonable: even if
in certain processes no poles occur, they will appear in the crossed channels.

As an example, the (expected) bounds $ h_3^Z \lesim 0.005 $ ,
$ h_4^Z \lesim 10^{-4} $ have been obtained for
the LHC~\refmark{\agbi} using the $ p \bar p \rightarrow Z \gamma \rightarrow
e^+ e^- \gamma $ reaction assuming that the CM energy
was $ 14 \tev $ while the scale of the form factor was $ 1.5 \tev $.
These values, however, imply that we have enough energy to
observe directly the heavy physics ($ 1.5 \ll 14 $).
The effective lagrangian approach breaks down in this region and
no reliable information can be derived from this approach, but this is
of little importance: the new physics would be directly observable.

Similar statements can be made for all form-factor modifications of
effective couplings. In fact there is an example from low-energy hadron
physics which illustrates the above statements. Consider the decay $ K
\rightarrow \pi e \nu $ which is characterized by two form
factors~\refmark{\okun} parametrized in the form $ 1 +
\lambda_{ K \pi }
q^2/ m_\pi^2 $; $q$ is the difference between the $K$ and $ \pi $
four-momenta, $ \lambda_{ K \pi } \sim 0.03 $ is a constant and $ m_\pi $ is
the
pion mass. This, to the same order in $q^2$ is equivalent to $ 1/( 1 -
q^2/ n M^2 )^n $, $ M \simeq 800 \gev $, which has poles at $ q^2 = n M^2
$. Of course we do find ``new'' physics (\ie\ physics beyond the
lightest pseudoscalar mesons) around $M$. It is also true that one
cannot simply replace the form factor by the expression $ 1/( 1 -
q^2/ n M^2 )^n $ in order to describe this new physics entirely.

\section{Large effects}

As I mentioned above, some operators are generated at tree level; for
the corresponding processes we do not expect an {\it a priori}
suppression. In this subsection I will give some examples. I will write
all operators in the unitary gauge.

\subsection{Fermion-gauge-boson couplings} Certain kinds of new physics can
induce {\it
right}-handed couplings of the $W$ to the quarks. The specific
interaction is $$ \inv{ \Lambda^2 } ( v + H )^2 \bar u_R \not \! W^+ d_R, $$
where $v=246\gev$ and $H$ denotes the Higgs field. Similar terms can be
generated for the $ b-t $ and the $ c-s $ quark pairs. Certain kinds of
physics also generate terms which mix generations.
Some such interactions are probed by the Michel-$\rho$ parameter, as well
as by the $W$ lifetime and branching ratios. All the bounds derived in
this way are relatively weak: $ \Lambda \gesim 500 \gev $ (for the first
and second generations only).

The couplings of the fermions to the $Z$ can be also modified. The
bounds derived from the oblique $\rho$ parameter as well as from the
LEP1 data are stronger than the ones above, implying $ \Lambda \gesim 2
\tev $~\refmark{\gw}.

\subsection{Higgs couplings} The presence of new interactions can modify
the couplings of the Higgs to the fermions, the Higgs self couplings and
the fermion-$V$-Higgs coupling ($V =W, \, Z $). An example of the latter
effects is the right handed current coupling described above.

\subsection{Four-fermion interactions} Many different kinds of physics
will generate four-fermion interactions, both CP violating and CP
conserving. Such interactions are strongly bounded if they occur between
first-family fermions. For the third family the bounds are generally
weak (or non-existent).

The operators generated by vector exchange have
been studied for NLC type of machines~\refmark{\ivf}. Similar
studies exist for LHC and other hadron colliders~\refmark{\ehlq}. I am
not aware of a comprehensive study (including scalar and vector exchange
possibilities) for the NLC.

\chapter{Strongly coupled theories}

When the underlying theory is strongly coupled~\foot{This possibly is usually
associated with the assumption that there is no light Higgs, I will
comment on this later.} the calculational reliability is
reduced for quantitative predictions.
 It is still possible, however, to
provide some reliable estimates~\refmark{\nda}. The idea is the following:
let $ \Lambda $ be the scale of new physics and assume that the
interactions of the particles lighter than $ \Lambda $ (the light
excitations) is described by some
effective theory which contains a series of
(effective) coupling constants. Just as in other theories, one can use the
effective theory to calculate the renormalization group evolution of
these couplings. In the case of strong coupling one must work to all
orders in perturbation theory, which is in general technically
impossible. One can, however, estimate the renormalization group
evolution and require that the running coupling constants do not diverge
at lower energies~\foot{In this argument it is assumed that no
cancellations occur between various graphs, \ie\ the theory is assumed
to be natural~\refmark{\nat}---no fine tunings are required.}.
This yields upper bounds on the various
coefficients of the terms in the effective lagrangian. It is interesting
to note that the {\it same} estimates for the $WWZ, \, WW\gamma$ and
$ZZ\gamma$ couplings derived above are obtained,
that is, the expressions
\estimatesi\ and \estimatesii\ are valid also for strongly coupled
underlying theories.

These arguments can be further specialized if it is assumed that there are no
light scalar particles, \ie\ that the low energy spectrum corresponds to
the Standard Model without the Higgs excitation. In this case the scale
of new physics $ \Lambda $ is constrained to be $ \lesim 4 \pi v \sim 3
\tev $~\refmark{\bound, \nda} and
some modifications occur which lead to refinements of the above bounds.
For example \estimatesi\ and \estimatesii\ are replaced by
$$ \lambda \sim { 6 M_w^2 g^2 \over 16 \pi^2 \Lambda^2} , \qquad
h_3^Z \sim { v^2 M_z^2 \over 16 \pi^2 \Lambda^4 } , \qquad h_4^Z \sim { M_z^4
\over
16 \pi^2 \Lambda^4 } , \qquad ( \Lambda\lesim 3 \tev) ; \eqn\eq $$
note that $h_{3,4}^Z$ acquire a strong
suppression factor, now a constraint $ | h_{3,4}^Z | < 1 $ imply $ \Lambda >
40 \gev, 26 \gev $ respectively.

\chapter{Rigidity of the bounds}

I have argued above that there is a way of estimating the couplings
which parametrize non-\sm\ effects using gauge-invariant effective
lagrangians. The question is then how rigid are these bounds.

Consider first the weakly coupled theory. One can argue that under
certain circumstances a given loop graph could be enhanced by having
several particles in the loop. This gives an order of magnitude leeway
in the above estimates (note that the same could be said about the
tree-level graphs).

What one cannot say is that there could be hundreds
of particles in the loop whose contributions cancel the $ 1 / (4 \pi )^2
$ entirely. If this were the case the theory would be such that the
one-loop effects would be as large as the tree-level ones and the theory
would be, in fact, strongly coupled. One can study such situations in
exactly-solvable toy models (J. Wudka, Ref. \reviews) and the result is that
the effect of this type of effects significantly alters the theory: it
is not possible (without significant fine-tuning) to maintain the scalar
mass below the cutoff. But, if the Higgs is no longer in the light
theory we must examine the model as a strongly coupled one. For this
case we revert to the arguments given in section 4 above.

For strongly coupled theories the bounds, as I mentioned, are more
qualitative. They are based on the assumption that no fine tuning should
be required of the underlying theory. If one grants this, the bounds
given hold (again with an order of magnitude uncertainty). I would also
like to add that these arguments can be applied to the non-linear sigma
model which describes low-energy hadron physics
and they agree well with experiment~\refmark{\gl}.

For example, allowing for a factor of $10$ enhancement in $ \lambda $
would imply that, in order to probe physics at the $1\tev$ level one
should measure $ \lambda $ to a precision of $ \lesim 10^{-3} $.
Similarly $ h_{3,4}^Z$ should be measured to a precision of $ \lesim 5
\times 10^{-3} $ and $ \lesim 7 \times 10^{-4} $ respectively.

While it is possible for some couplings to be thus enhanced, it is also
possible for them to be suppressed, either accidentally or as a result
of a symmetry. Thus a strong constraint on a given effective coupling
might indicate either a large value of $ \Lambda $ or the fact that the
underlying theory suppresses the coefficient under consideration. If
{\it all} effective couplings expected {\it a priori} not to contain the
(small) factors of $ 1/ ( 4 \pi ) $ are measured to be very small, the
simplest possibility is that this is a result of large $ \Lambda $, still all
possible scenarios should be considered when analyzing the data.

If one allows for fine tuning several of these statements can be
obviated. In this case, however, consistency would allow us to fine tune
anything we want, such models contain no information. Of course one
could say that Nature has chosen to fine tune just those interactions
which we have not probed directly, and while this is a logical
possibility, I will not consider it.

\chapter{Conclusion}

{}From the arguments given one can conclude that there is a
reliable method for extracting information about the scale of new
physics from the existing and expected data. In deriving the dependence
of the measurements on this scale one can work in a model independent
way using effective lagrangians. This does {\it not} mean, however, that
the coefficients can in principle have any values whatever: general
consistency requirements forbid their being too large and provide
estimates for their value. Using this input one can then determine the
reach into the realm of heavy physics that a given experiment has.

I have also strived to show that triple boson couplings are not the best
place to look for deviations from the standard model. Despite the fact
that one can write down lagrangians which appear to generate easily
observable deviations for these couplings, such ``models'' cannot be
derived from {\it any} consistent theory, weakly or strongly coupled,
with or without light scalars. The point is that one cannot state, by the mere
fact that a coefficient is measured to be small compared to one, that
the corresponding experiment is a sensitive probe of new physics. For
example, measuring the anomalous magnetic moment of the muon to $
10^{-7} $ says nothing about non-Standard Model physics~\refmark{\gmtwo}
(taking chiral symmetry to be natural~\refmark{\nat}).

The formalism presented determines the constraints an experiment should
satisfy in order to probe new physics up to a given scale.
For example in order for a $ 500 \gev $ collider to probe $WWZ$
physics beyond $ 1 \tev $, $ \lambda $ should be measured to a precision
better than $ \sim 10^{-4} $. Similar precision is required for the
parameter $ \Delta \kappa $ (related to the heavy physics contributions
to the $W$ magnetic and electric quadrupole moments~\refmark{\agbi}).

The processes which are worth measuring are those for which the
coefficients of the effective operators are as large as possible. It is
of course possible that the underlying theory will suppress precisely
those couplings, but I believe it is better to look at these terms than
to concentrate on terms which we are {\it certain} provide very small
effects.

The interactions with the largest coefficients have been catalogued for
the case of weakly-interacting heavy physics~\refmark{\tree}. It is also
possible to determine the kind of physics responsible for each of the
tree-level generated operators. Examples of such interactions are the
four-fermion interactions (generated by scalar of vector
exchanges~\foot{Tensor exchanges are reduced using Fierz
identities.}), and the $Z$ couplings to fermions; the particular case of
the $ Z b \bar b $ vertex can be shown to receive its largest
contributions through either $Z-Z'$ mixing (where $Z'$ denotes a heavy
vector boson) or through mixing of $b$ with some heavy fermions. Both
of these possibilities have been studied in the literature~\refmark{\zbb};
it can be
shown using the results of~\refmark{\tree} that these are the {\it only}
possibilities: no other kind of heavy physics can alter this vertex as
significantly.

Thus the effective lagrangian approach can not only estimate
coefficients, but can also exhibit the culprits responsible for any
operator. It is precisely the insistence that the underlying physics
should be described by a consistent model (whatever the details) that
imposes the
various constraints discussed above. If such consistency requirements
are foregone, the coefficients can indeed take any values, but in this
case the underlying physics is not described by {\it any} consistent
theory.

\ack
The author would like to thank M. Einhorn and K. Riles for many useful
comments. This work was supported in part through funds provided by the
Department of Energy

\refout

\bye